\documentclass[dvips]{article}
\usepackage{graphicx,amssymb,amsmath,times,subfigure}
\usepackage{icrctc07}

\title{Expected discovery potential and sensitivity of the ANTARES neutrino telescope to neutrino point-like sources.}
\shorttitle{Neutrino point-like sources in ANTARES}

\authors{J.~A. Aguilar$^{1}$ on behalf of the ANTARES collaboration}
\shortauthors{J.~A. Aguilar et al.}
\afiliations{$^1$IFIC -- Insituto de F\'{\i}sica Cospuscular,
	 CSIC -- Universitat de Val\`encia, apdo. 22085, E-46071 Valencia, Spain}
\email{aguilars@ific.uv.es}

\abstract{The ANTARES telescope is being built in the Mediterranean Sea. The
detector consists of a 3D array of photomultipliers (PMTs) that detects the
Cherenkov light induced by the muons produced in neutrino interactions. Other
signatures can also be detected. Since the neutrino fluxes from point-like
sources are expected to be small, it is of the utmost importance to take
advantage of the ANTARES pointing accuracy (angular resolution better than
0.3 degrees for muon events above 10 TeV) to disentangle a possible signal
from the unavoidable atmospheric neutrino background. In order to distinguish
an excess of neutrino events from the background, several searching
algorithms have been developed within the ANTARES collaboration. In this
contribution, the discovery potential and sensitivity to point-like sources
of the ANTARES neutrino telescope are presented.}

\begin{document}
\maketitle
\section{Introduction}

The ANTARES collaboration~\cite{ANTARES} has started the construction of an
underwater neutrino telescope in the Mediterranean Sea at a depth of
2475~m. Seven of the twelve lines of the detector have been deployed so far
(May 2007) and the whole detector will be commissioned in early 2008. Each
line is equipped with 75 10'' photomultiplier tubes (PMTs) joined in triplets
making 25 floors along the line. The lines are kept vertical by means of a
buoy located at their upper end. The mean distance among lines is about 65 m
and the instrumented length starts at 100 m over the seabed and covers about
350 m.
The angular resolution for the ANTARES telescope at high energies (above
10~TeV) is better than 0.3$^{\circ}$. At lower energies, the angular
resolution is dominated by the angle between the muon track and the original
neutrino direction.  The angular resolution together with the effective area
determine the performance of a neutrino telescope. The good pointing
accuracy and large effective area at high energies (0.06 km$^2$ at 100 TeV)
enables us to search for point-like neutrino sources or, in case no hint of
neutrinos source is found, to set restrictive upper limits to neutrino
fluxes.
\section{Search for point-like sources}
In the search for point-like steady sources, the whole visible sky
from the ANTARES location is surveyed making no assumption about time
correlation with observations of any other physical phenomenon, like GRBs,
flares and other transient sources. In this contribution, we will introduce a
new method for the search for point-like sources. This method is based on the
Expectation-Maximisation (EM) algorithm which is widely used as a likelihood
maximisation algorithm for clustering analysis. 
\subsection{The EM algorithm}

The Expectation-Maximisation~\cite{EM} algorithm is a general approach to
maximum-likelihood estimations for finite mixture model problems. For this method
a parametrisation of both the signal and neutrino background
density distributions is required. The main assumption is that sources signals are
supposed to follow Gaussian distributions which is reasonable assumption in
our case. The background distribution is inferred from the Monte Carlo data
sample, but it could be obtained directly from the real data by scrambling
the right ascension coordinate of the measured events. No energy information
or further performances of the detector are used in this case. The parameters
are determined maximising the likelihood using the EM algorithm. The EM
method assumes the existence of {\it missing} or {\it hidden} information,
in our case the knowledge whether a neutrino event comes from a given
source or it is produced by the atmospheric neutrino background. Hence, the
real observed data can be understood as an {\it incomplete} sample, so that,
adding a new vector {\bf z} we can build the {\it complete} data sample,
where {\bf z} is a class indicator vector that tells whether an event comes
from a source or not. Maximisation of likelihoods analytically intractable
can be easily accomplished by means of this methodology.  The EM general
method follows an iterative procedure where each iteration has two steps:

\begin{enumerate}

\item E-step: The expected value of the {\it complete} data log-likelihood,
  conditional to the observed data, is computed for a given set of parameters
  \{{$\boldsymbol{\Psi}^{(m)}$}\}
\item M-step: Find $\boldsymbol{\Psi} = \boldsymbol{\Psi}^{(m+1)}$ that
  maximises the expected value. This maximisation will lead to the
  maximisation of the desirable log-likelihood of the {\it complete} data
  sample  
\end{enumerate}
The model selection criteria is performed by means of the {\it Bayesian
  Information Criterion} or BIC. This BIC value is an approximation of the
  integrated likelihood when the number of events is high enough. The BIC
  value can be used as a test statistic and the discovery potential or
  discovery power of a point-like source to be detected can be inferred using
  the hypothesis testing theory. This discovery power will be described in
  section~\ref{sec: 4}.




\subsection{Other searching algorithms}
In ANTARES, other clustering algorithms have been developed in order to
exploit the potential of the detector in the search for point-like
sources. The first approach was a binning method used to look for a cluster
of events in different bins of a grid in which the sky is
divided~\cite{emi}. The significance of clusters is estimated comparing with
the distribution of atmospheric neutrino events which is uniform in right
ascension. These methods rely on the knowledge of the angular resolution of
the detector in order to build the optimum grid. Nevertheless, no hypothesis
is made on the neutrino source apart from the neutrino emission according to
a power law with spectral index of 2.

The first algorithm that did not rely on a binning approach was a Likelihood
Ratio method (LR)~\cite{aart}. This method operates by testing the data
compatibility with two hypotheses: one, often called {\it H$_{0}$}, is the
{\it null} hypothesis and the other one called {\it H$_{1}$}. The {\it null}
hypothesis is assumed to be the presence of background only, and {\it
H$_{1}$} considers the existence of a neutrino point source in addition to
the atmospheric background. The parametrisation of the density functions are
based on the obtained information from the simulated expected performances of
the detector.

\section{Foreseen results for the search for point-like sources in ANTARES}
\label{sec: 4}

\begin{figure}
\begin{center}
\noindent
\includegraphics [width=0.4\textwidth]{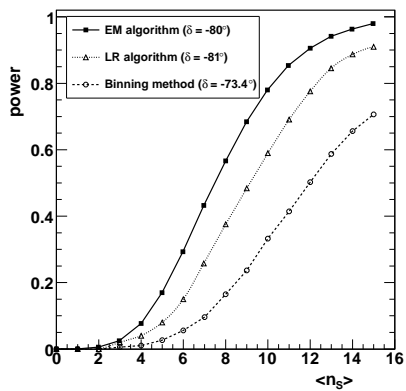}
\end{center}
\caption{Discovery power (probability of detection) for a source located at
  low declinations as a function of the averaged detected events for three
  methods developed in ANTARES for 5$\sigma$ confidence level.}\label{fig1}
\end{figure}

In this section, we will review the results of the point-like source search
in ANTARES based on a {\it blind search} over one year of data taking. A {\it
blind search} is called when no assumption about the source is done beyond
the educated assumption of a power law neutrino emission from the
sources. This type of search differs from the fixed point searches, where a
set of candidate sources are studied. The most direct way of presenting the
results in the searching analysis is by means of the discovery power, also
called discovery potential. This parameter accounts for the percentage of
success of discovering a point-like source over the atmospheric neutrino
background. Therefore given a number of events per year observed at the
detector from a source, the power of the method is the probability of
detecting such source with a given confidence level. This definition enables
the construction of plots as the one shown in figure~\ref{fig1}. This figure
shows the discovery potential for a confidence level of 5$\sigma$ as a
function of the mean number of observed events from a source located at the
vicinities of $\delta = -80^{\circ}$. The three main searching algorithms are
shown.
It is important to mention that the results yielded by the different analysis
correspond to different Monte Carlo simulations, muon track reconstruction
strategies, and therefore different performances of the detector. Hence a
direct comparison of the results obtained is not completely
fair. Nonetheless, we can observe that unbinned techniques result in a sound
improvement with respect to the classical binning approach. This result
motivates the use of unbinned techniques in the point-like analysis of real
data. Moreover, in this work, the energy estimate information is not used in
the EM algorithm which makes it specially appealing for the initial years of
operation in a neutrino telescope where the information from the muon energy
estimate does not constitute a solid piece of knowledge.

\begin{figure}
\begin{center}
\noindent
\includegraphics [width=0.4\textwidth]{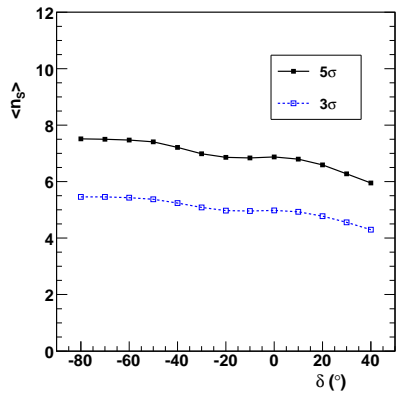}
\end{center}
\caption{Mean number of observed events required to yield a discovery power
  of 50\% for different confidence levels as a function of declination using
  the EM algorithm.}
\label{fig2}
\end{figure}

The discovery potential depends on the source location in the sky, specially
it depends only on the source declination since the atmospheric neutrino
background is right ascension independent. In order to show more clearly this
dependence, it is commonly presented as the required number of observed
events from a source to claim its existence in 50\% of equivalent experiments
as a function of the declination. Figure~\ref{fig2} shows this magnitude for
two different confidence levels using the EM algorithm. The required number of events is larger at lower declinations since the
number of atmospheric neutrinos per solid angle is larger at low
declinations.
In order to present the results in terms of a neutrino/muon flux, we can
divide the required number of events by the exposure of the detector in one
year to that declination. 
\begin{figure}
\begin{center}
\noindent
\includegraphics [width=0.4\textwidth]{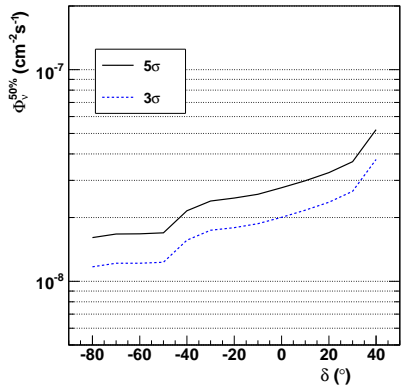}
\end{center}
\caption{Energy integrated (above 10~GeV) neutrino flux needed to claim the existence
  of a source with probability of 50\% for two confidence levels and the EM
  algorithm for a E$^{-2}$ spectrum.}
\label{fig3}
\end{figure}
Figure~\ref{fig3} shows the required integrated neutrino flux needed to claim
the existence of a neutrino source with a probability of 50\% as a function
of the declination. The required flux increases with declination mainly due
to the visibility factor. From the ANTARES location, lowest declinations are
visible 100\% of the time, whereas high declinations have a small visibility,
so that higher fluxes are needed to make a discovery.

\subsection{Sensitivity}

\begin{figure}
\begin{center}
\noindent
\includegraphics [width=0.4\textwidth]{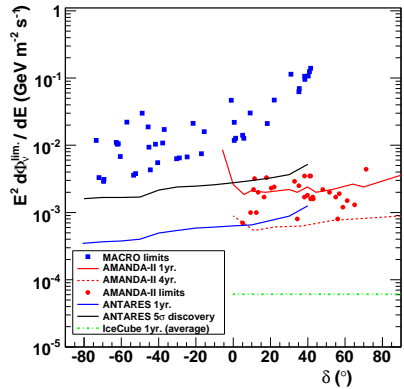}
\end{center}
\caption{Sensitivity of ANTARES in one year compared to other
  experiments. See text for further details.}
\label{fig4}
\end{figure}

If no excess of events is observed in the sky and therefore, no evidence of a
neutrino point-like source is claimed, we can set an upper limit on the
expected flux from a neutrino emitter. This upper limit can be provided in
terms of a full sky {\it blind search}. Nonetheless, most commonly it is
presented in terms of a point to point upper limit. Moreover, the concept of
upper limit is only applied when the experiment is currently running, an
averaged upper limit or sensitivity where the average upper limit is obtained
by an ensemble of $10^4$ experiments. Figure~\ref{fig4} shows the expected
sensitivity for ANTARES for one year of data taking compared to the published
results from other experiments. MACRO~\cite{MACRO} results are computed after
alive-time of 6.3 years. AMANDA upper limits are taken
from~\cite{AMANDA1YEAR,AMANDA4YEARS}. The projected sensitivity of IceCube
averaged over all Northern declinations is shown~\cite{ICECUBE}. The
required flux for a 5$\sigma$ discovery at 50\% for ANTARES in one year is also
indicated.

\section{Summary and conclusions}

In this contribution we have reviewed the expected performance of the
ANTARES neutrino telescope regarding the search for point-like neutrino
sources. 
Several searching algorithms have been devised in ANTARES. Among them,
unbinned techniques turn out to be more efficient than the standardised
binning approach. In addition, the method based on the EM algorithm presents
very good results without the trade-off of higher dependence on the estimated
detector performances. 
Discovery potentials in terms of number of events and required neutrino flux to claim the existence of a source at
50\% of probability has been presented. The comparison with other experiments
in term of the sensitivity has also been presented.
\\
\\
{\it This work is supported by Spanish MEC grants 
FPA2003-00531 and FPA2006-04277.}
\bibliography{icrc0638}
\bibliographystyle{unsrt}
\end{document}